\documentclass[preprint,aps]{revtex4}
\input boxedeps.tex
\SetRokickiEPSFSpecial  
\HideDisplacementBoxes

\newcommand{\be}{\begin{equation}}
\newcommand{\eq}{\end{equation}}
   
\begin{document}
\mbox{}\hfill SWAT/411\\
\vspace{10mm}
\begin{center}
{\Huge  Generalised Parton Distributions of the \\
\vspace{3mm} Pion on the Transverse Lattice${}^*$}\\
\vspace{20mm}
{\Large S. Dalley}\\ \vspace{5mm}
{\em Department of Physics, University of Wales Swansea, \\
Singleton Park, Swansea SA2 8PP, United Kingdom} \vspace{20mm}
\end{center}

The quark generalised parton distributions of the pion
are calculated from lightcone wavefunctions in  transverse lattice gauge 
theory at large $N_c$. The pion effective size is found to decrease with 
increasing momentum transfer. An analytic ansatz, consistent with finite 
boundstate lightcone energy conditions, is given  for the lightcone momentum 
dependence of the wavefunctions. This  leads to simple, universal
predictions for the behaviour of the distributions near the endpoints, 
complementing numerical DLCQ data. 

\vfil
${}^*$ Invited talk 
at {\em  Light-Cone 2004}, Amsterdam, Netherlands,  16-20 August, 2004. 
To appear in proceedings as a special issue of Few-Body Systems.
\newpage
\baselineskip .2in


\section{Introduction}
\label{intro}
The generalised parton distribution of the pion 
has  been
studied in a number of different models \cite{other}.
Following earlier work \cite{mat1}, van de Sande and the 
author \cite{dv1} recently computed numerically
the lightcone wavefunctions of the pion
in transverse lattice gauge theory at large $N_c$ \cite{bard1, review} (see
also \cite{chiral}). In this
talk, these wavefunctions will be applied to deduce the generalised
quark parton distributions in both  transverse momentum and position
space \cite{impact}. We also present an analytic ansatz for
the lightcone momentum dependence of the wavefunctions that leads
to simple, universal predictions for endpoint behaviour of the
distributions, complementing the numerical data.

In transverse lattice gauge theory in $3+1$ dimensions, 
two coordinates $x^\alpha$, $\alpha \in
\{0,3\}$, are continuous
while two directions ${\bf x}=\{x^1,x^2\}$ are discrete.
There are longitudinal continuum
gauge potentials $A_0, A_3$, transverse flux link fields $M_1, M_2$, and
fermions $\Psi$. 
For transverse lattice spacings $a$ of order the hadron size, 
the strategy is to perform
a colour-dieletric expansion \cite{dv2} of the
most general lightcone Hamiltonian, renormalisable with respect to
the continuum coordinates, in powers of $\Psi, M_1, M_2$ after eliminating
$A_0, A_3$ by lightcone gauge fixing. Provided the fields $\Psi, M_1, M_2$
are chosen sufficiently heavy, one can truncate this expansion to study the
low lying hadron boundstates dominated by just a few particles
of these fields. The remaining free couplings in the effective
Hamiltonian are fixed by optimizing symmetries broken by
the regulator and, if necessary, phenomenology.

We take the pion wavefunctions from Ref.~\cite{dv1}, which 
 contains details of the 
construction of the lightcone Hamiltonian and Fock space, the renormalisation
and the determination of various residual couplings appearing 
in the lightcone hamiltonian.
The couplings were constrained by 
optimizing covariance of low-lying meson and glueball wavefunctions,
rotational invariance of the heavy source potential, and fitting 
two phenomenological
parameters in the meson sector, 
conveniently taken to be $f_\pi$ and the $\rho$ meson mass. In addition,
two fundamanetal scales are taken as given, the $\pi$ meson mass and the 
string tension $\sqrt{\sigma}$. 

A typical large-$N_c$ 
meson eigenstate state consists of linear combinations of 
gauge invariant 
basis states formed from quarks separated by a chain of transverse
links $\overline{\Psi} MM\cdots M \Psi$.
If we
label the transverse orientation of link fields
by indices $\lambda_j \in \{\pm 1, \pm 2\}$,
the partonic decomposition for a meson of momentum
$P=(P^-,P^+,{\bf P})$ when the transverse momentum ${\bf P} = 0$
can be written
\begin{eqnarray}
|P^+, {\bf P} & = & {\bf 0} \rangle = \sum_{n=2}^{\infty} \int [dx]_{n}
\sum_{{\bf z}_1, {\bf z}_n, h, h^{\prime}}
\left[\sum_{\lambda_j}\right]_{n}
\psi_n(x_{i}, h, h^{\prime}, \lambda_{j}) \times \nonumber \\
&& |(x_{1}, h, {\bf z}_{1}); 
(x_2, \lambda_1, {\bf z}_{2}) ); \cdots  
  ; (x_{n-1}, \lambda_{n-2}, {\bf z}_{n-1}); 
(x_{n}, h^{\prime}, {\bf z}_{n})  \rangle \ , 
\label{decom}
\end{eqnarray}
where $h, h^{\prime}$ denote quark and anti-quark helicities respectively,
${\bf z}_i$ is the transverse position and
$x_i$ the fraction of $P^+$ carried by the $i^{\rm th}$ parton.
We have
\be 
\int [dx]_{n}  =  \int dx_1 \cdots dx_n \delta\left( \sum_{i=1}^{n}
 x_i - 1\right) 
\eq
and $\left[\sum_{\lambda_j}\right]_{n}$ indicates that the sum over
orientations of links must form an unbroken chain on the transverse lattice
between quark and anti-quark.
We define a set of hadron states boosted to general transverse 
momentum ${\bf P}$
by applying
the Poincar\'e generators ${\bf M}^{+} = (M^{+1}, M^{+2})$. 
This gives for each parton Fock state
\begin{eqnarray} 
&& {\rm exp}\left[ -{\rm i} {\bf M}^{+}. {\bf P}/P^+ \right]
|(x_{1}, h, {\bf z}_{1}); \cdots ; (x_{n}, h^{\prime}, {\bf z}_{n})  
\rangle \nonumber \\
& = &
{\rm exp}\left[ {\rm i} {\bf P}. {\bf c} \right]
|(x_{1}, h, {\bf z}_{1}); \cdots  ; 
(x_{n}, h^{\prime}, {\bf z}_{n})  \rangle \ , \label{boost}
\end{eqnarray}
\be
{\bf c} = (c^1 , c^2) = \sum_{i=1}^{n} x_i {\bf z}_{i} \ .
\eq

The lightcone wavefunctions $\psi_n(x_{i}, h, h^{\prime}, \lambda_{j})$
computed in Ref.\cite{dv1} were subject to a number of cutoffs.
Firstly, 
the transverse lattice spacing is held fixed at 
$a \approx 300 {\rm MeV}^{-1}$, representing the renormalisation
scale. Thus, it would be inappropriate to use momentum transfers $Q$
much larger than this. 
A cutoff $n\leq 5$ was used. This translates into a maximum
transverse size for the meson of 3 links, or about 2~fm. The effect
is to reduce slightly the charge radius of the pion from its experimental
value, as was shown in
the form factor $F(Q^2)$ calculated in Ref.\cite{impact}.
Finally, calculations were performed in DLCQ, which discretizes
the momentum fractions $x_i$ in units of a small cutoff $1/K$, to create
a finite-dimensional Fock space. The largest basis
studied was $K=20$, which is of dimension 82,470. Observables tend not
to change much when extrapolated from $K=20$ to $K = \infty$, except
close to the edges of phase space.
We will give an analytic approach to these edges in section~\ref{analytic}

\section{Numerical Generalised Parton Distribution}
\label{dlcq}
We follow the conventions of Diehl \cite{diehl}, defining the generalised
distributions in lightcone gauge as
\begin{equation}
{\cal H}(\overline{x},\xi,Q^2)=
{1 \over \sqrt{1-\xi^2}} \int_{-\infty}^{+\infty} {dz^- \over 4 \pi}
{\rm e}^{{\rm i} \overline{x} \overline{P}^{+} z^-}
\langle P_{\rm in} | \overline{\Psi}(-z^-/2) \gamma^+
\Psi({z}^-/2) | P_{\rm out} \rangle \ ,
\label{gpd}
\end{equation}
\begin{eqnarray}
Q & = &  P_{\rm in} - P_{\rm out} \\ 
\xi & = & {( P_{\rm in} - P_{\rm out})^+ \over 2\overline{P}^+} \\
\overline{P}^+ & = & {(P_{\rm in}  + P_{\rm out})^+ \over 2}
\end{eqnarray}

\begin{figure}
\centering
$\displaystyle {\cal H}(x,0,Q^2)$\hspace{5pt}
\BoxedEPSF{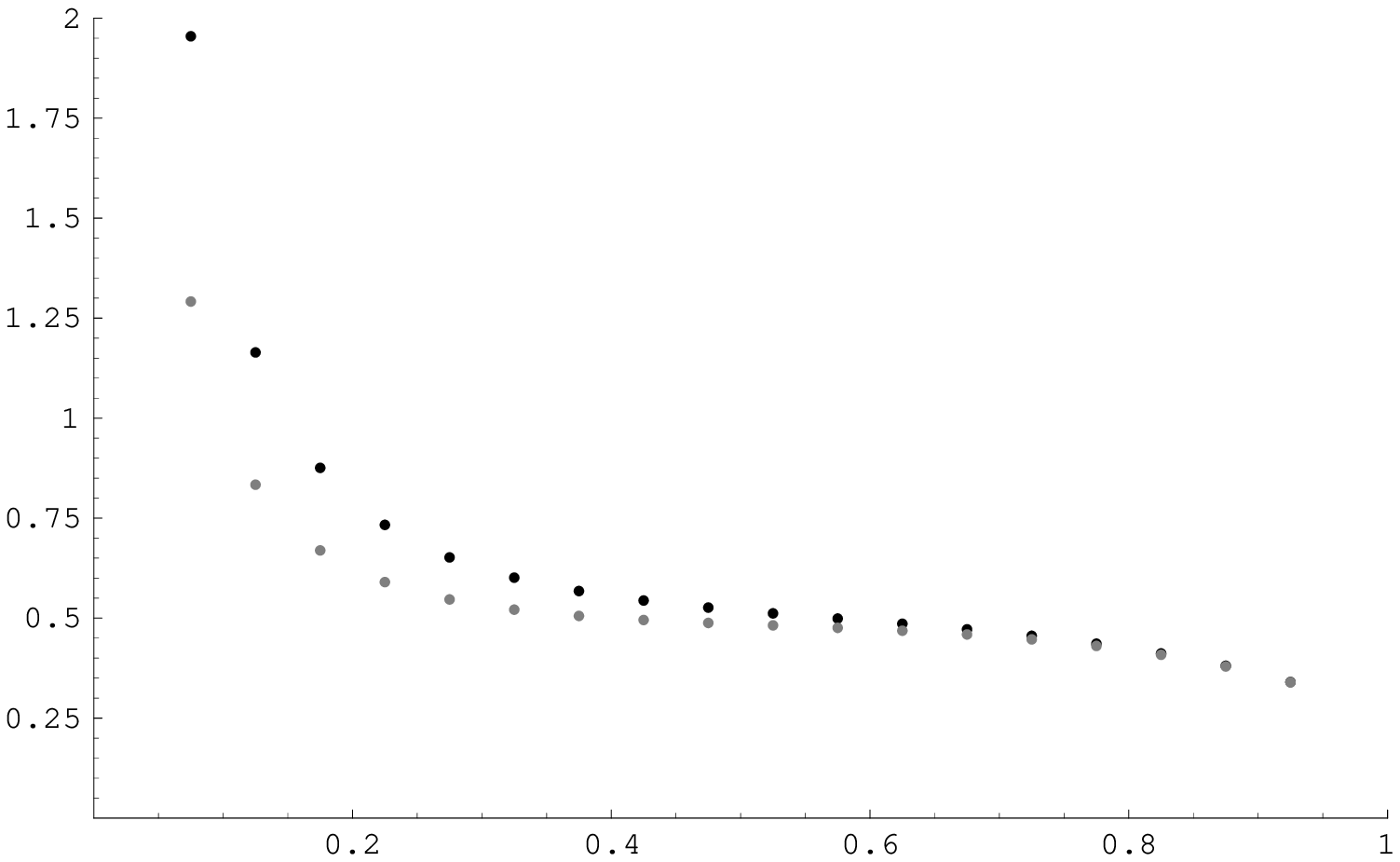 scaled 725}\\
\hspace{0.5in}$\displaystyle x$\\
\caption{
Generalised quark distribution of the pion at zero skewdness and
DLCQ cutoff $K=20$: black points are $Q=0$, grey are $Q=500\ {\rm MeV}$.
\label{fig1}}
\end{figure}

Unfortunately, we are not able to explore the region $\overline{x} < \xi$
with the strict large-$N_c$ wavefunctions. That region is governed by
their $1/N_c$ corrections, since it depends upon quark pair production.
Even at $\overline{x} > \xi$, in DLCQ one needs 
to form overlaps of wavefunctions with
different $K$ in fixed ratio for fixed $\xi$. Existing data do not cover
enough values of $K$. Thus, we concentrate on $\xi=0$ for the DLCQ analysis.

The $K=20$ data are displayed in fig.1 for 
two extreme (!) values of $Q$ at $\xi =0$. 
${\cal H}(x,0,0)$ is the conventional structure
function \cite{dv1}, which exhibits rising Regge behaviour at small $x$
characteristic of a system with gluonic degrees of freedom. It falls
to zero precipitously near $x \to 1$. This  may be an indication that 
the chiral limit is near (the pion mass was fit to $140 \ {\rm MeV}$), since
chirally symmetric quark models \cite{ruiz1}  produce
a flat distribution. As purely transverse $Q$ is increased, the distribution
is seen to deplete at small $x$ but is $Q$-independent as $x \to 1$.

\begin{figure}
\centering
$\displaystyle B^2 (Q^2)$\hspace{5pt}
\BoxedEPSF{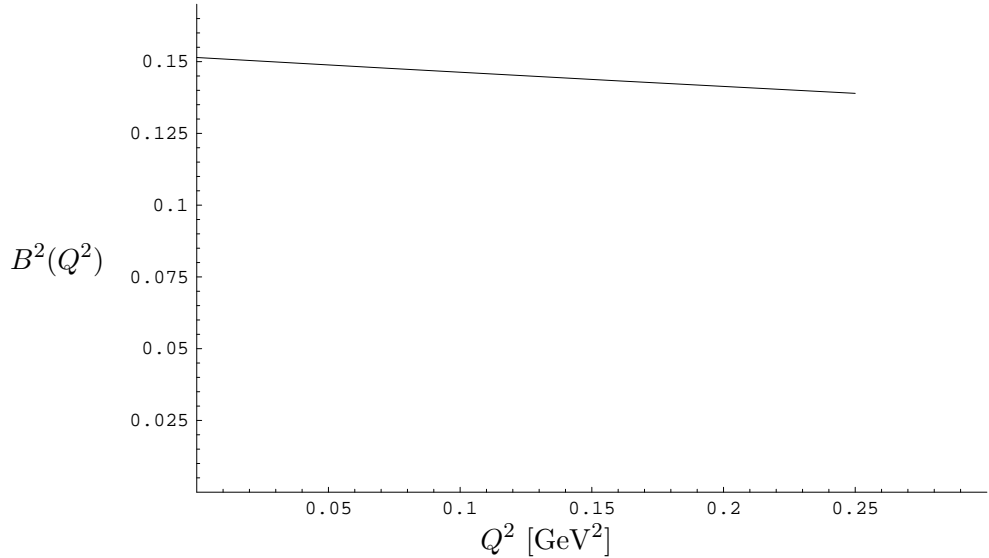 scaled 750}\\
\hspace{0.5in}$\displaystyle Q^2 \ [{\rm GeV}^2]$\\
\caption{
Effective size of the pion as measured by Eq.~(\ref{size}).
\label{fig2}}
\end{figure}
A measure of the effective size of the pion is given by the average
distance between the active quark and the center of momentum of the
spectators \cite{mat2}
\begin{equation}
B^2 (Q^2) = - {1 \over F(Q^2)}{\partial \over \partial Q^2}
\int_{0}^{1} dx \ {{\cal H}(x,0,Q^2) \over (1-x)^2} \ .
\label{size}
\end{equation}
This is plotted in fig.~2. The authors of Ref.~\cite{mat2} claim
that $B^2(Q^2 \to \infty) \to 0$ is a condition for colour
transparency. The data are consistent with the pion size reducing, but
obviously it is impossible to reliably extrapolate into  the large
momentum transfer region.

\section{Impact Parameter Distribution}

The impact parameter dependent quark distribution was analysed in 
Ref.\cite{impact}. It
is defined by \cite{diehl2}
\begin{eqnarray}
{\cal I}(\overline{x},\xi,{\bf b})& = &{1 \over \sqrt{1-\xi^2}}
\langle {P}_{\rm out}^{+}, {\bf b}^{\rm out}_{0} | 
\int {dz^- \over 4 \pi} e^{{\rm i} \overline{x} \overline{P}^+ z^-} 
 \times \nonumber \\
&& \hspace{5mm} \overline{\Psi}
(-z^-/2, {\bf b})
\gamma^+ \Psi(z^-/2, {\bf b})
|{P}_{\rm in}^{+}, {\bf b}^{\rm in}_{0} \rangle \ , \label{skew}
\end{eqnarray}
where 
\be
{\bf b}^{\rm out}_{0}=-{\xi \over 1-\xi} {\bf b} \ \ , \ \ 
{\bf b}^{\rm in}_{0}={\xi \over 1+\xi} {\bf b} \ .
\eq
${\bf b}_{0}$ is the transverse position 
of the hadron, assumed spinless. 
${\cal I}(x,0,{\bf b})$ is simply the probability
of a quark carrying fraction $x$ of the lightcone momentum $P^+$ when 
at transverse position
${\bf b}$ \cite{mat3}.

\begin{figure}
\centering
\vspace{-50mm}
\BoxedEPSF{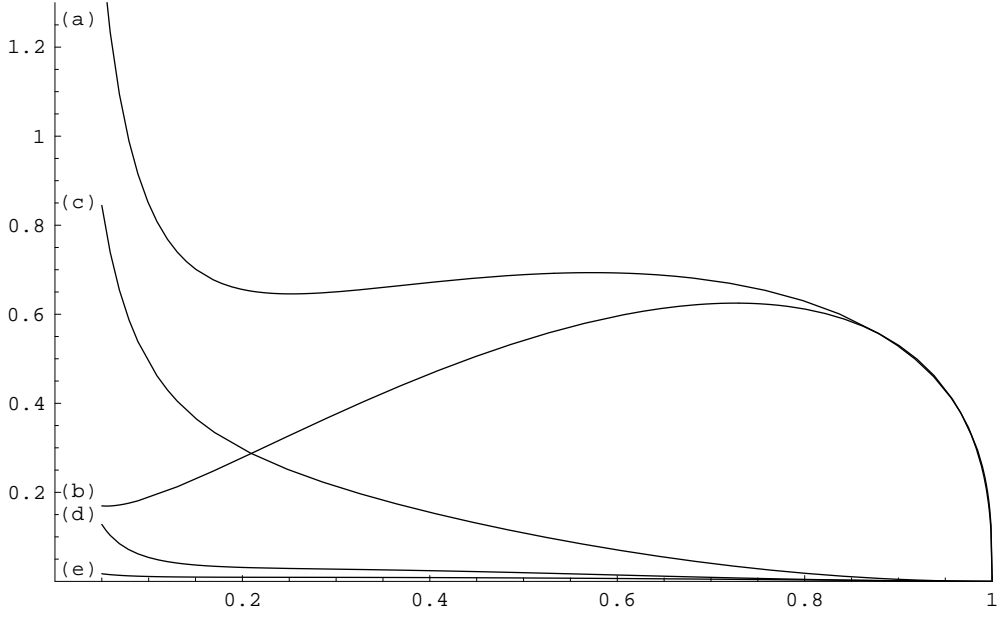 scaled 800}\\
\vspace{-50mm}
\caption{Impact parameter dependent valence quark distributions of the
pion. (a) The full distribution ${\cal H}(x,0,0)$, which  sums 
over all impact parameters ${\bf b}$. (b)-(e) The sum of contributions 
at impact parameters ${\bf b}$ along  lattice axes such that 
$b= | {\bf b}| =0,a,2a,3a$ respectively. 
\label{fig3}}
\end{figure}

To translate results to hadron impact 
parameter space in the transverse directions, we 
integrate ${\bf P}$ over the Brillouin zone to obtain a hadron
state localised at transverse position ${\bf 0}$
\begin{eqnarray}
|P^+, {\bf b}_{0} ={\bf 0} \rangle & = & 
\sum_{n=2}^{\infty} \int [dx]_{n}
\sum_{{\bf z}_1, {\bf z}_n, h, h^{\prime}}
\left[\sum_{\lambda_j}\right]_{n}
{a^2 \over \pi^2} {\sin{c^{1}\pi/a} \over c^1}{\sin{c^{2}\pi/a} \over c^2}
\times \nonumber \\ 
&& \psi_n(x_{i}, h, h^{\prime}, \lambda_{j}) 
|(x_{1}, h, {\bf z}_{1}); \cdots  ; (x_{n}, h^{\prime}, {\bf z}_{n}) 
\rangle
\end{eqnarray}
Hence
\begin{eqnarray}
{\cal I}(x,0,{\bf b}) &= &\sum_{n=2}^{\infty}
\int [dx]_{n}\sum_{{\bf z}_1, {\bf z}_n, h, h^{\prime}}\left[\sum_{\lambda_j}\right]_{n} \delta(x_1 -x) \delta({\bf z}_1 -
{\bf b}) \times \nonumber \\
&&\left[{a^2 \over \pi^2} {\sin{c^{1}\pi/a} \over c_1}
{\sin{c^{2}\pi/a} \over c_2}\right]^2
|\psi_n(x_{i}, h, h^{\prime}, \lambda_{j})|^2 \ . 
\end{eqnarray}
Due to the lattice cutoff, this should strictly be interpreted
as the probability of finding the quark within one lattice spacing
of impact parameter ${\bf b}$. We show results for the first few
values of $b=|{\bf b}|$ sampled along lattice axes in fig.~3. In this case,
the curves are the result of an extrapolation to $K = \infty$ from
data at $K=10,12,15,20$.  Fig.~3(a) shows the full distribution ${\cal H}(x,0,0)$, 
which  sums 
over all impact parameters ${\bf b}$ (one can compare this with the
$K=20$ version in Fig.~1).
Fig. 3(b) is the contribution from ${\bf b} = (0,0)$, Fig. 3(c)
is the sum of contributions from ${\bf b} = (a,0),(0,a),(-a,0),(0,-a)$,
etc..
This clearly
shows that the valence region $x > 0.5$ is dominated by quarks
with impact parameter less than the charge radius $\sim 2/3$~fm.
It also suggests that the rapid rise at small $x$ is dominated
by larger impact parameters $b > 2/3$~fm. 
This all implies a
sharp fall-off of the hadron wavefunction in transverse space at a 
particular ($x$-dependent) radius.

\section{Analytic Wavefunction Ansatz}
\label{analytic}
The boundstate problem can be cast into a set of integral equations for the
wavefunctions $\psi_n$
\begin{equation}
M^2 \psi_n(x_{i}, h, h^{\prime}, \lambda_{j}) = \int {\cal K} \cdot \psi_n
\label{bound}
\end{equation}
where, on the right hand side,
 the wavefunction is convoluted with a kernel ${\cal K}$
\cite{dv1}
that exchanges lightcone momentum, creates, and annihilates partons.
Typically the kernel is singular when one or more momentum
fractions $x_i$ vanish in this equation. Demanding that the boundstate mass $M$
remains finite, one can constrain the behaviour of the wavefunctions
$\psi_n$ in these limits \cite{ladder}. A thorough analysis of the endpoint
behaviour of transverse lattice wavefunctions will be given in 
Ref.~\cite{endpoint}. Here we will give the simplest ansatz for the
first few $\psi_n$ in a large-$N_c$ meson that 
has the correct endpoint behaviour when any number of $x_i$ vanish:
\begin{eqnarray}
\psi_2(x) & = & C_2 x^{\alpha} (1-x)^{\alpha} \\
\psi_3(x,y) & = & C_3 {y^{\beta} (1-y)^{\alpha}\over
[(x+y)(1-x)]^{\beta-\alpha+1/2}} \\
\psi_{4}(x,y,z) & =&   C_4  {(yz)^{\beta} [(1-y)(1-z)]^{\alpha} 
[(x+y+z)(1-x)]^{\alpha+\beta-1/2}
\over  [(x+ y)(1-x-y)]^{\beta  +1/2}(y+z)^{2\beta}  } \\
& \ldots & \nonumber 
\label{ansatz}
\end{eqnarray}
The exponents $\alpha$ and $\beta$ are independent of $n$, quark helicity,
and the transverse shape, while
the coefficients $C_n$ are dependent on all of these factors.
They are all determined by the solution to
Eqs.~(\ref{bound}). The coefficients $C_n$ are complex in general, although
overlaps of wavefunctions appearing in ${\cal H}$ are real because
the theory is $PT$ invariant.
For normalisability one must have $\beta < 1/2$.
The simplest forms above
could be generalised for use as a variational basis,
by distinguishing quark and link-field $x_i$ and/or multiplying them by
a complete basis of analytic functions of the $x_i$.
In principle, one can use DLCQ data to determine the unknown
parameters but, in practice, it is difficult because convergence is slow
in the endpoint regions. We could, however, obtain a fairly reliable
fit $\alpha \approx 1/2$ from the data of Ref.~\cite{dv1}, implying
a quark distribution function of the form $\sqrt{x(1-x)}$ at this low
scale.

Even without determining the unknown parameters, one can deduce a number
of interesting conclusions. Firstly, one notes that the diverging singularity
of the wavefunctions $\psi_n$ becomes stronger as  more momentum
fractions vanish. Intuition from
the free lightcone kinetic energy $\sim  \sum_i 
m_i /x_i$, for partons of mass $m_i$,
would have suggested the complete opposite. Rather than simply vanishing
at the edges of phase space, as indicated by the free theory,
the interactions force 
relations between wavefunctions with different numbers of partons to cancel
the divergences in Eq.~(\ref{bound}). This was noted for continuum QCD
for a single vanishing $x_i$ in the first reference of Ref.\cite{ladder},
but for the transverse lattice we are able to generalise it to any
number of vanishing momenta. This allows us to make a number of
definite statements relevant to generalised quark parton distributions:
\begin{itemize}
\item[1.]
each $\psi_n$ contributes ${\rm const.}\times  (1-\overline{x})^{2 \alpha}$ to
${\cal H}(\overline{x},\xi, Q^2)$ as $\overline{x} \to 1$;
\item[2.]
$\psi_n$ contributes $(\ln {1 \over x})^{n-3}$, $n > 2$,  to
${\cal H}({x},0,0)$ as ${x} \to 0$;
\item[3.]
each $\psi_n$, $n >2$, contributes a  finite constant to 
${\cal H}(\overline{x},\xi, Q^2)$ as $\overline{x} \to \xi$ from above;
\item[4.]
$\partial {\cal H}({x},0, Q^2)/\partial Q^2 |_{Q^2 = 0} \sim (1-x)^2$
as $x \to 1$.
\end{itemize}
Provided the constants fall sufficiently fast with $n$, continuity 
\cite{diehl} and 
item 3 above suggest that ${\cal H}(\xi,\xi,Q^2)$ is finite.

\section{Outlook}
The transverse lattice has so far proven to be the only systematically
improvable non-perturbative lightcone method that preserves the
gauge invariance crucial to confinement.
The current DLCQ transverse lattice calculations for mesons can be improved 
by going to higher orders of the colour dielectric expansion and testing
directly for $P$, $T$, and chiral symmetry. As well as the initial 
applications to the pion presented here, one could  also calculate
the generalised distributions of the heavier mesons. 
Of course, most
experimental data derive from the nucleon, so it is important to 
develop computations in the baryon sector. These have proven awkward
in the past because the numerical work does 
not simplify in the large-$N_c$ limit. However,
many of the analytic arguments above will also be valid for baryons.

\vspace{10mm} 
\noindent Acknowledgements: 
This work is supported by PPARC grant 
PPA/G/0/2002/00470.  I thank Ben Bakker, Daniel Boer, and Piet Mulders for
their work to organise this workshop.

\end{document}